\documentclass{ws-ijmpe}
\begin{document}

\markboth{J. Nordhaus}{A Binary Scenario for the Formation of Strongly Magnetized White Dwarfs}

\catchline{}{}{}{}{}

\title{A Binary Scenario for the Formation of Strongly Magnetized White Dwarfs}

\author{\footnotesize J. NORDHAUS\footnote{NSF Astronomy and Astrophysics Postdoctoral Fellow} \footnote{nordhaus@astro.rit.edu}}

\address{Center for Computational Relativity and Gravitation\\
Rochester Institute of Technology\\
Rochester, NY 14623, USA\\
\&\\
National Technical Institute for the Deaf\\
Rochester Institute of Technology\\
Rochester, NY 14623, USA\\
\&\\
Dept. of Astrophysical Sciences\\
Princeton University\\
Princeton, NJ 08544, USA}

\maketitle

\begin{history}
\received{(received date)}
\revised{(revised date)}
\end{history}

\begin{abstract}
Since their initial discovery, the origin of isolated white dwarfs (WDs) with magnetic fields in excess of $\sim$1~MG has remained a mystery.  Recently, the formation of these high-field magnetic WDs has been observationally linked to strong binary interactions incurred during post-main-sequence evolution.  Planetary, brown dwarf or stellar companions located within a few AU of main-sequence stars may become engulfed during the primary's expansion off the main sequence.  Sufficiently low-mass companions in-spiral inside a common envelope until they are tidally shredded near the natal white dwarf.  Formation of an accretion disk from the disrupted companion provides a source of turbulence and shear which act to amplify magnetic fields and transport them to the WD surface.  We show that these disk-generated fields explain the observed range of magnetic field strengths for isolated, high-field magnetic WDs.  Additionally, we discuss a high-mass binary analogue which generates a strongly-magnetized WD core inside a pre-collapse, massive star.  Subsequent core-collapse to a neutron star may produce a magnetar.  
\end{abstract}

\section{Observational Constraints}

Approximately 10\% of isolated white dwarfs form with magnetic field strengths in excess of $10^6$ G\cite{1}.  Based on Zeeman measurements, surface field strengths for these high-field magnetic white dwarfs (HFMWDs) range from a few to slightly less than a thousand megagauss (MG)\cite{2,3,4}.  This is in contrast to the bulk of isolated white dwarfs with measured weak fields or non-detection upper limits of typically $\lesssim10^4-10^5$~G\cite{5,6,7}.

If the magnetic field strengths of white dwarfs were independent of binary interactions, then the observed distribution of isolated WDs should be similar to those in detached binaries.  Remarkably, {\it not a single} detached, binary system in which the primary is a WD and the companion is an M-dwarf contains a HFMWD\cite{8,9,10}.  Within 20 pc, there are 109 known WDs (21 of which have a non-degenerate companion).  SDSS has identified 149 HFMWDs ({\it none} of which has a non-degenerate companion).  The maximum probability of obtaining samples at least this different from the same underlying population, assuming binomial statistics\cite{11}, is $5.7\times10^{-10}$.  This suggests at the 6.2-$\sigma$ level that the two populations are different.  Furthermore, SDSS identified 1253 WD+M-dwarf binaries ({\it none} of which are magnetic).  As was previously suggested, the presence or absence of binarity is crucial in influencing whether a HFMWD results\cite{12}.  These results initially seem to indicate that HFMWDs preferentially form when isolated.  However, unless there is a mechanism by which very distant companions prevent the formation of a strong magnetic field, a more natural explanation is that highly-magnetized white dwarfs became that way by engulfing (and removing) their companions\cite{13}. 

\section{The Binary Scenario}

In the binary scenario, the progenitors of HFMWDs are those systems that incur a common envelope (CE) phase during post-MS evolution.  Magnetic CVs may be the progeny of common envelope systems that almost merge but eject the envelope, leaving a close binary.  Isolated HFMWDs may result from systems with less massive companions -- as suggested by recent observations.  In this case, the companion can't eject the envelope and thus, continues to in-spiral until it tidally disrupts.  The formation of an accretion disk can both amplify magnetic fields and transport them to the WD surface.  Recently, this scenario was suggested and quantitatively investigated as a way to produce an isolated HFMWD\cite{13}.  Figure 1 sketches the evolutionary pathway to forming an isolated HFMWD.  A close, M-dwarf+main-sequence binary will enter a common envelope when the primary leaves the main-sequence either through direct expansion or tidal engulfment\cite{19}.  The M-dwarf continues to in-spiral toward the natal WD core until it tidally disrupts and forms an accretion disk.  The magnetic field is then amplified in the accretion disk and transported onto the WD surface.  Mass-loss proceeds during the post-main-sequence-giant phases until the natal, magnetized WD is exposed.  

\begin{figure}[th]
\centerline{\psfig{file=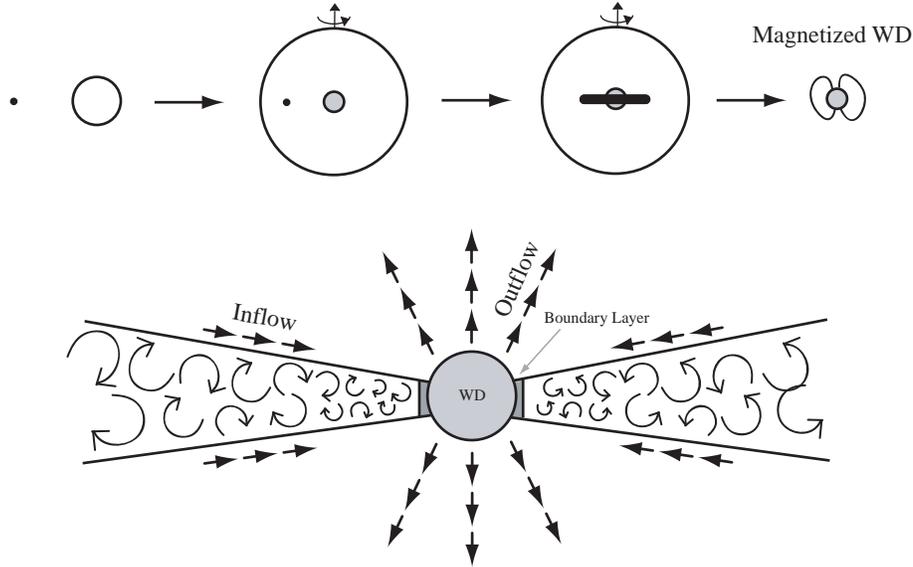,width=12cm}}
\vspace*{8pt}
\caption{A path to form an isolated, highly magnetized WD.  The top panel shows an evolutionary sequence.  From left to right: A close, main-sequence binary $\longrightarrow$ common envelope phase during the post-main sequence $\longrightarrow$ tidal disruption and formation of an accretion disk $\longrightarrow$ emergence of a HFMWD.  Note that although the cartoon depicts a dipole field, the final magnetic geometry (small scale field vs. large scale field) will depend on details of the amplification process.  The bottom panel shows the turbulent accretion disk.  A conveyor belt of magnetized material flows in.  Some material is lost to outflows while some is incorporated into the white dwarf.  Additional field amplification may occur in the boundary layer between the slowly rotating WD and Keplerian accretion disk.}
\end{figure}

In this proceeding, we highlight recent results from this scenario\cite{13} and show that for a range of disrupted-companion masses, the fields generated in the disk are sufficient to explain the observed range of HFMWDs.  Additionally, we discuss the possible evolutionary relationships that HFMWDs have with related objects such as post-AGB stars\cite{14,15} and planetary nebulae (PNe)\cite{16,17,18,19}.  Finally, we comment on high-mass stellar analogues\cite{13}.

\subsection{Disk Formation}
\label{ssec:DiskDynamo}

Disks can form inside common envelopes for a wide range of companions.  Jupiter-mass planets\cite{20}, brown dwarfs and low-mass stars will tidally disrupt inside an AGB star when the self gravity of the companion is overcome by the differential gravitational force of the white dwarf core.  This occurs at the tidal shredding radius, which we estimate as 
\begin{equation}
R_{\rm s} \simeq R_{\rm c} (2M / M_{\rm c})^{1/3},
\end{equation} 
where $r_{\rm c}$ and $M_{\rm c}$ are the radius and mass of the companion and $M$ is the stellar mass interior to $R_{\rm s}$ \cite{19}.
Note that planetary and low-mass stellar companions to solar-type stars are significantly more plentiful than brown dwarfs\cite{21}.  Therefore, disks in the lower and higher mass ranges may be more common than those in the intermediate mass range.

Once the disk forms, it is ionized and thus, susceptible to the development of magnetized
turbulence (e.g.~via the magneto-rotational instability).  Accretion proceeds on a viscous timescale given by $t_{\rm visc} \simeq R^{2}/\nu \simeq P_{\rm orb} / \alpha_{\rm ss} (H / R)^2$,
where $P_{\rm orb}$ is the Keplerian orbital period at radius $R$, $\nu$ is the effective kinematic viscosity, $H$
is the disk scaleheight, $c_{\rm s} = H\Omega$ is the midplane soundspeed, 
$\Omega = 2\pi / P_{\rm orb}$ is the Keplerian orbital frequency, and $\alpha_{\rm ss}$ is the
dimensionsless Shakura-Sunyaev parameter that characterizes the
efficiency of angular momentum transport.  Initially, the accretion rate can be approximated as
\begin{eqnarray}
\dot{M} &\sim& \frac{M_{\rm c}}{t_{\rm visc}|_{r_{\rm s}}} \approx 7M_{\odot}{\,\rm yr^{-1}}\times \label{eq:mdot0} \\
&& \left(\frac{\alpha_{\rm ss}}{10^{-2}}\right)\left(\frac{M_{\rm c}}{30 M_{\rm J}}\right)^{3/2}\left(\frac{r_{\rm c}}{r_{\rm J}}\right)^{-1/2}\left(\frac{H/R}{0.5}\right)^{2} \nonumber \, ,
\end{eqnarray}
where $r_{\rm c}$ is scaled to the radius of Jupiter and $H/R$ to a large value $\sim 0.5$ because the disk cannot is geometrically thick and cannot cool efficiently at high accretion rates.  Additionally, since the companions under consideration lead to disks much more dense than the stellar envelope, we ignore any interaction between the disk and the star.

Equation (\ref{eq:mdot0}) illustrates that for $M_{\rm c} \sim
0.1-500 M_{\rm J}$, and for typical values of $\alpha_{\rm ss}
\sim 0.01-0.1$, $\dot{M}$ is initially a few to many orders of magnitude larger
than the Eddington accretion rate of the proto-WD (i.e., $\dot{M}_{\rm Edd} \sim 10^{-5} M_{\odot}$~yr$^{-1}$).  At first, it might seem apparent that inflow of mass onto the WD surface would be
inhibited by radiation pressure.  However, this
neglects the fact that, at sufficiently high accretion rates, photons
are trapped and advected to small radii faster than they can diffuse
out \cite{22,23,24}.  In this `hyper-critical' regime,
accretion is possible when $\dot{M} \gg \dot{M}_{\rm Edd}$.

Hyper-critical accretion may occur if the accretion timescale, $t_{\rm visc}$, is less than the timescale for photons to diffuse out of the disk midplane, $t_{\rm diff}$\cite{23}.  The location in the disk where this occurs is called the ``trapping radius'' and is given by $R_{\rm tr} = \dot{M}\kappa H / 4\pi R c$ where $\kappa$ is the opacity.  An important quantity is the ratio of the trapping radius to the outer
disk radius $R_{\rm s}$ (coincident with the tidal shredding radius) :
\begin{eqnarray}
\frac{R_{\rm tr}}{R_{\rm s}} &\approx& 1.0\times 10^{4}\left(\frac{\alpha_{\rm ss}}{10^{-2}}\right)\left(\frac{\kappa}{\kappa_{\rm es}}\right)\times\label{eq:trap_ratio} \\
&& \left(\frac{M_{\rm c}}{30 M_{\rm J}}\right)^{11/6}\left(\frac{M_{\rm WD}}{0.6M_{\odot}}\right)^{-1/3}\left(\frac{r_{\rm c}}{r_{\rm J}}\right)^{-3/2}\left(\frac{H/R}{0.5}\right)^{3}, \, \nonumber
\end{eqnarray}
where we have used equation (\ref{eq:mdot0}) and scaled $\kappa$ to
the electron scattering opacity $\kappa_{\rm es} = 0.4$ cm$^{2}$ g$^{-1}$.  If $R_{\rm s}<R_{\rm tr}$, then $t_{\rm visc}<t_{\rm diff}$ and photons are advected with the matter.

From equation (\ref{eq:trap_ratio}) we conclude that $R_{\rm tr}> R_{\rm s}$ for $M_{\rm c} \gtrsim 0.1M_{\rm J}$.  This implies that photon pressure will be unable to halt accretion initially and that the local energy released by accretion must be removed by advection\cite{25}.  Advection acts like a conveyor belt, nominally carrying the gas to small radii as its angular momentum is removed.  If there is no sink for the hot gas, this conveyor may ``jam".  This is an important distinction between white dwarfs and neutron stars or black holes, as the latter two can remove the thermal energy by neutrino cooling or advection through the event horizon, respectively.  For WDs, neutrino cooling is ineffective and thermal pressure builds, such that radiation pressure may again become dynamically significant.  However, this scenario neglects the possibility of outflows, which can sustain inflow by carrying away the majority of the thermal energy, thereby allowing a smaller fraction of the material to accrete at a higher than Eddington rate.  Though more work is needed to assess the efficiency of outflows in the present context, radiatively-inefficient accretion flows are widely thought to be prone to powerful outflows\cite{26,27}.  Even if accretion occurs at, or near, the Eddington rate, the fields produced ($\gtrsim10$~MG) are still strong enough to explain the bulk of the magnetized WD population.  The origin of the strongest field systems ($\sim$100-1000 MG) may be problematic if accretion is limited to the Eddington rate.

Nevertheless, the material deposited outside the WD will be hot and virialized, forming an extended envelope with a lengthscale comparable to the radius.  Though hot, this material will eventually cool on stellar timescales and become incorporated into the stellar layers near the proto-WD surface.  If this material cools at some fraction of the local Eddington luminosity, it will be incorporated onto the WD surface on a timescale $\sim$$10^2-10^4$ years -- much less than typical AGB lifetimes.

\subsection{Disk Dynamo}
Due to the presence of shear in the disk, the MRI is a likely source of turbulence,
and can supply the conditions by which magnetic field amplification can occur.  Large-scale fields produced by the MRI have been modeled via $\alpha-\Omega$ dynamos at various levels of sophistication.  However, for the purposes of estimating orders of magnitude of the fields, approximate values from less sophisticated treatments can be employed as we do now.  The Alfv\'{e}n velocity in the disk obeys $v_{\rm
  a}^2=\alpha_{\rm ss} c_{\rm s}^2$, such that
the mean toroidal field at radius $R$ is given by\cite{28}
\begin{eqnarray}
\label{eq::toroidal}
\overline{B}_\phi &\sim& \left(\frac{\dot{M}\Omega}{R}\frac{R}{H}\right)^{1/2}  \\
&\approx& 160~\rm{MG}\left(\frac{\eta_{\rm acc}}{0.1}\right)^{1/2}\left(\frac{\alpha_{\rm ss}}{10^{-2}}\right)^{1/2}\left(\frac{M_{\rm c}}{30 M_{\rm J}}\right)^{3/4}\times \nonumber \\
&&\left(\frac{M_{\rm WD}}{0.6M_{\odot}}\right)^{1/4}\left(\frac{r_{\rm c}}{r_{\rm J}}\right)^{-1/4}\left(\frac{H/R}{0.5}\right)^{1/2}\left(\frac{R}{10^{9}\rm cm}\right)^{-3/4} \, , \nonumber
\end{eqnarray}
where in the second equality we have substituted equation
(\ref{eq:mdot0}) for $\dot{M}$, and
multiplied by the factor $0.1 \lesssim \eta_{\rm acc} \le 1$ to
account for the possibility of outflows as described above.  If an
$\alpha-\Omega$ dynamo operates, the mean poloidal field
$\overline{B}_p$ is related to the toroidal field via
$\overline{B}_p=\alpha_{\rm ss}^{1/2}\overline{B}_\phi$\cite{28}.  However, regardless of whether a large scale
field is generated, a turbulent field of magnitude $\overline{B}_{\phi}$ is likely to be present and contributing significantly to the Maxwell stresses responsible for disk accretion.

The toroidal field evaluated near the WD
surface $R \approx R_{\rm WD} \approx 10^{9}$~cm as a function of
companion mass $M_{\rm c}$, calculated for two different values of the
viscosity ($\alpha_{\rm ss} = 0.01$ and $0.1$) is shown in Fig. 2.  In both cases, we assume that $\eta_{\rm acc} = 0.1$.\\

{\it Note that for the range of relevant companion masses $M_{\rm c} \sim 0.1 - 500$
$M_{\rm J}$, $\overline{B}_{\phi} \sim 10-1000$~MG, in precisely the
correct range to explain the inferred surface field strengths of
HFMWDs (see Fig. 2 below).}

\begin{figure}[th]
\centerline{\psfig{file=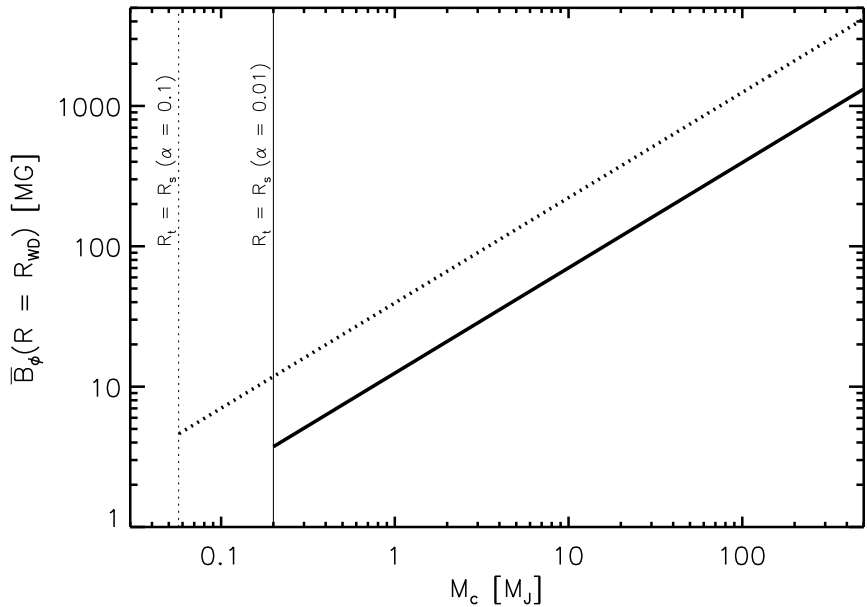,width=12cm}}
\vspace*{8pt}
\caption{Toroidal magnetic field strength,$\overline{B}_{\phi}$ (eq.~[\ref{eq::toroidal}]), at the WD surface
   as a function of the mass of the tidally disrupted companion $M_{\rm c}$.  Toroidal field strengths are presented
  for two values of the viscosity, $\alpha_{\rm ss} = 0.01$ ({\it
    solid line}) and $\alpha_{\rm ss} = 0.1$ ({\it dotted line}) and
  assuming an accretion efficiency $\eta_{\rm acc} = 0.1$.  The
  white dwarf mass and radius are $M_{\rm WD} = 0.6M_{\odot}$ and
  $r_{\rm WD} = 10^{9}$~cm, respectively.  The vertical lines show the
  companion mass above which photons are trapped in the accretion flow
   (i.e., $r_{\rm tr} > r_{\rm s}$;
  eq.~[\ref{eq:trap_ratio}]), such that super-Eddington accretion
  occurs.
}
\end{figure}

\section{Proto-Planetary and Bipolar Planetary Nebulae}

If the origin of highly magnetized WDs lie in common envelope phases, then they may also exist around proto-planetary (PPNe) and planetary nebulae (PNe).  For these systems, the link between close binary interactions and the formation of PPNe/PNe has been well established\cite{14,15,29}.  It's interesting to also note that the presence of outflows inside the AGB star, near the proto-WD surface is already coincident with the prevalence of outflows in proto-planetary nebulae.  The direct detection of a HFMWD in a PPN or PN system would provide additional evidence for the binary scenario.  Recent, indirect evidence in the form of hot X-ray emitting gas in two post-CE PN systems has been discovered and could be due to magnetic activity or accretion onto the WD\cite{30}.  Additionally, observations of masers in post-AGB water fountain sources show magnetically-shaped, bipolar jet-like structures\cite{31,32}.

\section{Higher Mass Stellar Analogues and Magnetars}

While $\sim10$\% of WDs are highly magnetized, it's interesting to note that $\sim10$\% of neutron stars are highly magnetized.  High-mass stars may also undergo common envelope interactions in the
presence of close companions.  If the common envelope field mechanisms
described in this proceeding and in our previous paper\cite{13} operate in high-mass stars, then the result could
be strong field neutron stars or magnetars (neutron stars with
magnetic fields in excess of $\sim$$10^{14}-10^{15}$~G).  In
particular, formation of an accretion disk from an engulfed companion
during a red supergiant phase could produce a magnetized WD core.  In
the eventual core-collapse and stellar supernova explosion (possibly
driven by the neutrino mechanism; \cite{33,34}), the
magnetized WD core collapses to a neutron star.  If simple flux
freezing operates (itself an open question) and the initial magnetized
core is on the order of $\sim$100-1000~MG, homologous collapse to a
neutron star would generate $\sim$$10^{14}-10^{15}$~G fields.  Typical
neutron stars that possess modest field strengths may originate from
core-collapse supernova of single stars or stars without having
incurred a CE phase.  Note that what we are proposing is an alternative to the neutrino-driven convection dynamo\cite{35}.  In our scenario, the engulfment of a companion and the formation of an accretion disk naturally provides fast rotation, magnetized turbulence and differential rotation -- the same ingredients that are also needed for the neutrino-driven convection dynamo.

\section{Conclusions}

In summary, common envelope evolution as the origin of
highly-magnetized, white dwarfs is strongly supported by observations.
For HFMWDs, whether the scenario described in this proceeding and our previous paper\cite{13} is ultimately found to be viable will depend on the statistics of low-mass stellar and substellar companions\cite{36} to
stars of similar masses to (or somewhat higher masses than) the Sun.
The numerous radial velocity searches of the last 20 years
have revealed a number of such companions though the precise occurrence rate of
companions in orbits that could lead to the kind of disk-dynamo mechanism remains unclear.  Further theoretical
work into the binary origin of HFMWDs and magnetars requires the
development of multi-dimensional, magnetohydrodynamic simulations of
the CE phase.

\section*{Acknowledgements}

JN is supported by an NSF Astronomy and Astrophysics Postdoctoral Fellowship under award AST-1102738 and by NASA HST grant AR-12146.01-A.


\begin{thebibliography}{0}
%
\bibitem{1} J.~Liebert, P.~Bergeron, \& J.~B.~Hoberg {\it Astronomical Journal} {\bf 125} (2003) 348.
%
\bibitem{2} B.~T.~Gansicke, F.~Euchner, \& S.~Jordan {\it Astronomy and Astrophysics} {\bf 394} (2002) 957.
%
\bibitem{3} G.~D.~Schmidt et al.~{\it Astrophysical Journal} {\bf 595} (2003) 1101.
%
\bibitem{4} K.~M.~Vanlandingham et al.~{\it Astronomical Journal} {\bf 130} (2005) 734.
%
\bibitem{5} A.~Kawka, S.~Vennes, G.~D.~Schmidt, D.~T.~Wickramasinghe \& R.~Koch {\it Astrophysical Journal} {\bf 654} (2007) 499.
%
\bibitem{6} G.~Valyavin et al.~{\it Astrophysical Journal} {\bf 648} (2006) 559.
%
\bibitem{7} R.~Aznar Cuadrado et al.~{\it Astronomy and Astrophysics} {\bf 423} (2004) 1081.
%
\bibitem{8} J.~Liebert et al.~{\it Astronomical Journal} {\bf 129} (2005) 2376.
%
\bibitem{9} N.~M.~Silvestri et al.~{\it Astronomical Journal} {\bf 131} (2006) 1674.
%
\bibitem{10} N.~M.~Silvestri et al.~{\it Astronomical Journal} {\bf 134} (2007) 741.
%
\bibitem{11} D.~S.~Spiegel, F.~Paerels \& C.~A.~Scharf {\it Astrophysical Journal} {\bf 658} (2007) 288.
%
\bibitem{12} C.~A.~Tout et al.~{\it MNRAS} {\bf 387} (2008) 897.
%
\bibitem{13} J.~Nordhaus, S.~Wellons, D.~S.~Spiegel, B.~D.~Metzger, E.~G.~Blackman {\it Proceedings of the National Academy of Science} {\bf 108} (2011) 3135.
 %
 \bibitem{14} J.~Nordhaus \& E.~G.~Blackman {\it MNRAS} {\bf 370} (2006) 2004.
 %
 \bibitem{15} J.~Nordhaus, E.~G.~Blackman \& A.~Frank {\it MNRAS} {\bf 376} (2007) 599.
 %
 \bibitem{16} B.~Miszalski et al.~{\it MNRAS} {\bf 413} (2011) 1264.
 %
 \bibitem{17} J.~Nordhaus et al.~{\it Astrophysical Journal Letters} {\bf 684} (2008) L29.
 %
 \bibitem{18} B.~Miszalski et al.~{\it Astronomy and Astrophysics Letters} {\bf 488} (2008) L79.
 %
 \bibitem{19} J.~Nordhaus et al.~{\it MNRAS} {\bf 408} (2010) 631.
 %
 \bibitem{20} D.~S. Spiegel, A.~Burrows \& J.~Milson {\it Astrophysical Journal} {\bf 727} (2011) 57.
 %
 \bibitem{21} J.~Farihi, E.~E.~Becklin \& B.~Zuckerman {\it Astrophysical Journal Supplements} {\bf 161} (2005) 394. 
 %
 \bibitem{22} S.~A.~Colgate {\it Astrophysical Journal} {\bf 163} (1971) 221.
 %
 \bibitem{23} J.~M.~Blondin {\it Astrophysical Journal} {\bf 308} (1986) 755.
 %
 \bibitem{24} R.~A.~Chevalier {\it Astrophysical Journal} {\bf 346} (1989) 847.
 %
 \bibitem{25} H.~C.~Spruit {\it The Neutron Star -- Black Hole Connection} Eds. C.~Kouveliotou, J.~Ventura, \& E.~van den Heuvel (2001) 141.
 %
 \bibitem{26} J.~Hawley \& S.~A.~Balbus {\it Astrophysical Journal} {\bf 573} (2002) 738.
 %
 \bibitem{27} K.~Ohsuga, M.~Mori, T.~Nakamoto \& S.~Mineshige {\it Astrophysical Journal} {\bf 628} (2005) 368.
 %
 \bibitem{28} E.~Blackman, A.~Frank \& C.~Welch {\it Astrophysical Journal} {\bf 546} (2001) 288.
 %
 \bibitem{29} O.~De Marco {\it PASP} {\bf 878} (2009) 316.
 %
 \bibitem{30} R.~Montez, O.~De Marco, J.~H.~Kastner \& Y.~Chu {\it Astrophysical Journal} {\bf 721} (2010) 1820.
 %
 \bibitem{31} N.~Amiri, W.~Vlemmings, H.~J.~van Lagenvelde {\it A\&A} {\bf 532} (2011) 149.
 %
 \bibitem{32} W.~Vlemmings \& H.~J.~van Lagenvelde {\it A\&A} {\bf 488} (2008) 619.
 %
 \bibitem{33} J.~Nordhaus, A.~Burrows, A.~Almgren, J.~B.~Bell {\it Astrophysical Journal} {\bf 720} (2010) 694.
 %
 \bibitem{34} J.~Nordhaus et al.~{\it Physical Review D} {\bf 82} (2010) 103016.
 %
 \bibitem{35} R.~C.~Duncan \& C.~Thompson {\it Astrophysical Journal Letters} {\bf 392} (1992) 9.
 %
 \bibitem{36} E.~Bear \& N.~Soker {\it MNRAS} {\bf 411} (2011) 1792.
 %
\end{thebibliography}
\end{document}